\documentclass[fleqn,usenatbib]{mnras}

\usepackage{newtxtext,newtxmath}
\usepackage[T1]{fontenc}
\usepackage{ae,aecompl}

\usepackage{graphicx}	% Including figure files
\usepackage{amsmath}	% Advanced maths commands
\usepackage{changes}    % to add changes in the text
\definechangesauthor[name=Gio, color=blue]{gio}
\usepackage{ulem}

\usepackage{soul}	% Pour barrer du texte
\usepackage{xcolor}
\newcommand{\beq}{\begin{equation}}
\newcommand{\eeq}{\end{equation}}
\newcommand{\bea}{\begin{eqnarray}}
\newcommand{\eea}{\end{eqnarray}}
\newcommand{\nn}{\nonumber}

\newcommand{\erre}{{\cal R}}
\newcommand{\Dif}{{\cal D}}
\usepackage{color}
\usepackage{subfigure}
\usepackage{multirow}
\usepackage{float}
\usepackage{nameref}
\usepackage{natbib}
\usepackage{soul}	
\usepackage{xcolor}

\graphicspath{{./figures/}}

\title[]{Different spectra of cosmic ray H, He and heavier nuclei escaping compact star clusters}

\author[]{
	Pasquale Blasi$^{1,2}$\thanks{E-mail: pasquale.blasi@gssi.it} \&
	Giovanni Morlino $^{3}$\thanks{E-mail: giovanni.morlino@inaf.it}
	\\
	$^{1}$ Gran Sasso Science Institute, Viale F. Crispi 7 - 67100 L' Aquila, Italy\\
	$^{2}$ INFN-Laboratori Nazionali del Gran Sasso, Via G. Acitelli 22, Assergi (AQ), Italy \\
	$^{3}$ INAF/Osservatorio Astrofisico di Arcetri, Largo E. Fermi, 5 - 50125 Firenze, Italy
}

\date{}

\pubyear{2023}

\begin{document}

\label{firstpage}
\pagerange{\pageref{firstpage}--\pageref{lastpage}}
\maketitle

\begin{abstract}
Cosmic ray acceleration at the termination shock of compact star clusters has recently received much attention, mainly because of the detection of gamma ray emission from some of such astrophysical sources. Here we focus on the acceleration of nuclei at the termination shock and we investigate the role played by proton energy losses and spallation reactions of nuclei, especially downstream of the shock. We show that for a reasonable choice of the mean gas density in the cavity excavated by the cluster wind, dominated by the presence of dense clouds, the spectrum of He nuclei  escaping the bubble is systematically harder than the spectrum of hydrogen, in a manner that appears to be qualitatively consistent with the observed and yet unexplained phenomenon of discrepant hardening. We also find that,  in this scenario, the spallation reactions of heavier nuclei are likely to be so severe that their spectra become very hard and with a low normalization, meaning that it is unlikely that heavy nuclei escaping star clusters can provide a sizeable contribution to the spectrum of cosmic rays at the Earth. Limitations and implications of this scenario are discussed.
\end{abstract}


%%%%%%%%%%%%%%%%%%%%%%%%%%%%%%%%%%%%%%%%%%%%%%%%%%

%%%%%%%%%%%%%%%%% BODY OF PAPER %%%%%%%%%%%%%%%%%%

%%% INTRODUCTION %%%
\section{Introduction}
\label{sec:introduction}

Star clusters have been invoked as potential sources of Galactic cosmic rays (CRs) mainly in connection with two issues: 1) the difficulties encountered by supernova remnants (SNRs) in accelerating particles to the rigidity of the knee \cite[]{Lagage1,Lagage2,Schure-Bell:2013,pierre}, and 2) the observed anomalous $^{22}$Ne/$^{20}$Ne abundance in CRs \citep{Gupta+2020}, which is a factor $\sim 5$ larger than for the solar wind \citep{Binns+:2006}, a result that is difficult to accommodate in the framework of particle acceleration at SNR shocks alone \citep{Prantzos2012}.

Recent observations of gamma ray emission from some stellar clusters (SC), such as Westerlund 1 \citep{Abramowski_Wd1:2012}, Westerlund 2 \citep{Yang+2018}, Cygnus cocoon \citep{Ackermann:2011p3159,Aharonian+2019NatAs}, NGC 3603 \citep{Saha_NGC3603:2020}, BDS2003 \citep{HAWC:2021}, W40 \citep{Sun_W40:2020} and 30 Doradus in the LMC \citep{HESS-30Dor:2015}, have revived the interest in these astronomical objects as potential sites of particle acceleration, a topic that has been more or less popular many times in the past \citep{Casse-Paul:1980, Cesarsky-Montmerle:1983,webb1985,Gupta+2018,Bykov+2020}. In particular, the detection of gamma rays with energy up to $\sim 100$ TeV with HAWC \cite[]{2021NatureHAWC} and the detection of photons beyond 1 PeV from the Cygnus cocoon by LHAASO \cite[]{LAHHSO2021Nature594} has boosted the investigation of these sources as accelerators of cosmic rays (CRs) up to the knee.  

Where exactly particle acceleration takes place in SCs is still unclear. One possibility is inside the cluster core, where CR energy gets boosted by crossing multiple shocks created by the collision of stellar winds \citep{Reimer:2006} or by efficient scattering by magnetic turbulence generated in such collisions \citep{Bykov:2001p3166} \cite[see also][for a review of different acceleration models]{Bykov+2020}. More recently \cite{Morlino+2021} proposed that particle acceleration could take place at the termination shock of the cluster wind, generated by the impact of the wind with the interstellar medium (ISM). Such a structure can indeed be formed if the SC is compact enough \citep{Gupta+2020}. 
Evidence that at least some fraction of CRs should be accelerated from stellar wind material comes from the study of CR chemical composition \citep{Tatischeff+2021}, strengthening the idea that stellar winds are important. A further possibility is that the acceleration is driven by individual SNRs exploding inside SCs \citep{Vieu+2022, Vieu-Reville:2023} with a possible additional contribution due to multiple shock crossing in case of overlapping SNRs \citep{Parizot:2004p3162, Ferrand-Marcowith:2010}. Such scenarios can be applied only to SCs older than $\sim 3$\,Myr, when the most massive stars start exploding.

In this work we will focus on the model of particle acceleration at the collective wind termination shock of relatively young SCs. A formal solution of the transport equation has been found by \cite{Morlino+2021} while a numerical solution of the same problem, including energy losses of the accelerated particles in the cavity excavated by the wind, has been developed by
 \cite{BlasiMorlino2023} and applied to the description of the gamma ray emission from the Cygnus OB2 association. The measurement of the spectrum and spatial distribution of the gamma ray emission from the Cygnus cocoon \cite[]{2021NatureHAWC} has provided us with a unique test bench for this theoretical framework. In particular, \cite{BlasiMorlino2023} made explicit predictions in terms of acceleration efficiency at the termination shock and interaction of CRs in the region where the gamma ray emission is produced, mainly as a result of neutral pion production and decay. The theoretical framework allows us to assess in a quantitative manner the possibility to reach PeV energies in star clusters and to determine the spectrum of particles escaping the bubble excavated by the collective wind of the stars. The latter is significantly affected by inelastic energy losses suffered by protons downstream of the shock to the extent that the mean gas density downstream of the termination shock, as due mainly to the presence of dense clouds of gas in that region, is $\gtrsim 10~\rm cm^{-3}$. 

Here we discuss in detail the acceleration and transport of CR nuclei inside the bubble excavated by the winds of the stars in the cluster. After acceleration at the termination shock, nuclei are advected and diffuse in the downstream region, before escaping the bubble. Energy losses and spallation reactions, especially in the downstream region, play a crucial role in shaping the spectrum of both protons and nuclei and affect what eventually escapes the bubble excavated by the star cluster wind. 

We show that for values of the mean density in the bubble that are comparable with those required to describe the observed gamma ray emission from selected clusters, the net effect is that of a hardening of the spectrum of He nuclei compared to that of protons, a phenomenon resembling the discrepant hardening, a term first coined by the CREAM collaboration \cite[]{2010CREAM}. The spectrum of H and He nuclei has now been measured by PAMELA \cite[]{PAMELA-P-He}, AMS-02 \cite[]{ams-P,AMS-He}, DAMPE \cite[]{DAMPE-P,DAMPE-He}, CALET \cite[]{CALET-P,CALET-He} and all agree that, after accounting for CR transport in the Galaxy, the source spectrum of helium nuclei is required to be harder than the spectrum of protons. This effect has been discussed in detail in \cite{Evoli1}. 
In fact, more recently, the AMS-02 measurements have shown that the spectra of nuclei heavier than He \cite[]{AMS-nuclei} are also required to be injected with spectra harder than that of protons \cite[]{Evoli1,Evoli2,Schroer2021Nuclei}. 
This finding appears to be at odds with what we know about diffusive shock acceleration, the main candidate mechanism for particle energization, that predicts equal spectra for nuclei at the same rigidity. While this is certainly the case in SNRs, where energy losses inside the acceleration region are negligible, as discussed above this conclusion does not necessarily apply to star clusters. 

One should notice that the spectrum of He measured at Earth includes the contribution of both $^3$He and $^4$He, hence the coupled transport equations for both nuclei need to be solved. Spallation of $^4$He is rather complex in that it mostly results in the production of $^3$He, which in turn is subject to spallation reactions. When those effects are accounted for in the transport through the Galaxy, the total spectrum of He nuclei does not show any hardening compared with that of protons \cite[]{Vladimirov2012}. Here we show that in the case of star clusters the total spectrum of He ($\rm ^{3}He+^{4}He$) remains harder than that of H, to an extent that depends on the mean gas density in the bubble. The spectral hardening is much more pronounced for heavier nuclei because of the larger cross sections, and for a gas density $\gtrsim 10 ~\rm cm^{-3}$, heavier nuclei show such a severe suppression for energies $\lesssim 100$ TeV to make us conclude that star clusters can only contribute a fraction of the observed flux of heavy nuclei at Earth. 

A crucial role in reaching these conclusions is played by the spatial distribution of cold (atomic and molecular) gas in the bubble: such gas represents the target for hadronic interactions that lead to the production of gamma rays, but is also the target for the spallation reactions of nuclei. In the case of Cygnus OB2 region, the average gas density estimated through the molecular line emission along the line of sight is of the order of $5\div 10\,\rm cm^{-3}$ \cite[]{menchiariPhD:2023}. Such density is mainly contributed by dense atomic and/or molecular gas clouds, filling a fraction of the volume of the cavity. While the total flux of gamma rays due to inelastic CR collisions is proportional to the mass in the form of clouds \cite[]{BlasiMorlino2023}, the effect of spallation, as discussed throughout this article, is related to the fraction of the time necessary to leave the bubble that cosmic rays spend inside clumps rather than in the dilute wind. We find that for typical density of an individual cloud of $n_{\rm cl} \sim (10^2-10^3)~\rm cm^{-3}$, the grammage accumulated by particles propagating downstream of the termination shock may become comparable with the critical grammage necessary for spallation of a nucleus.

The clumps of material invoked here may be clouds that  have survived the wind of the SC or, to some extent, be the result of the fragmentation of the dense shell of material ploughed away by the wind during its expansion, as shown by hydrodynamic (HD) simulations \cite[]{Lancaster+2021a, Lancaster+2021b}\footnote{Notice, however, that the simulations by \cite{Lancaster+2021a, Lancaster+2021b} adopt values of the wind power and density of circumstellar medium which are not representative of the most massive SCs such as, for instance, Cygnus OB2. Moreover, those simulations do not account for the presence of magnetic fields which, in general, tend to suppress HD instabilities.}.
As discussed by \cite{BlasiMorlino2023}, the clumping of the cold gas in clouds may have some effect on the structure of the bubble through the processes of cooling and evaporation. More important, the processes of gamma ray production and nuclear spallation may turn out to be dependent upon the ability of CRs to penetrate the neutral gas in the clouds. So, although the case in favor of a scenario in which the gamma ray production is predominantly of hadronic origin and nuclei suffer serious spallation in the bubble seems rather compelling, the exact extent of these phenomena are bound to depend on details of the structure of the bubble that we do not have at our disposal at present. 

The article is organised as follows: in \S \ref{sec:bubble} we briefly summarize the general properties of the wind blown cavity. In \S~\ref{sec:transport} we describe the solution of the transport equation of CR nuclei in the cavity and the associated diffusive particle acceleration at the termination shock. In \S \ref{sec:results} we illustrate our results both in terms of hardening of the spectrum of He escaping the bubble and in terms of spallation of heavier elements. In \S \ref{sec:concl} we outline our conclusions.
 
%%% SECTION %%%
\section{The wind blown bubble}
\label{sec:bubble}

A detailed discussion of the mechanisms involved in the formation of a cavity blown by the collective winds of the stars in a cluster was provided in \cite{Morlino+2021}, hence here we will limit ourselves with a short summary of the relevant quantities. We assume here to be dealing with a compact star cluster, namely a cluster in which the termination shock is located well outside the region where the stars are concentrated \cite[]{Gupta+2020}. The wind due to the collective effect of all stars has velocity $v_{w}$ and density:
\beq
\rho_w(r) = \frac{\dot M}{4\pi r^{2} v_{w}},~~~ r > R_{c},
\eeq
where $R_{c}$ is the radius of the core where the stars are concentrated, and $\dot M$ is the rate of mass loss due to the collective wind. 

The interaction of this wind with the ISM surrounding the cluster results in the formation of a weak forward shock that moves slowly outwards in time, and a strong termination shock where the wind is slowed down and heated up, that can be considered as quasi-stationary.

At first sight, since the typical cooling timescale of the shocked ISM is only $\sim 10^4$ yr, while the cooling time for the shocked wind is several $10^7$ yr \citep{Koo-McKee:1992a, Koo-McKee:1992b}, one can safely assume that the wind-blown bubble evolves quasi-adiabatically. However, as discussed by \cite{BlasiMorlino2023}, the presence of clumps of cold gas in the bubble may affect the goodness of this assumption, through cooling and evaporation of the clumps. \cite{Gupta+2016} retained the effect of cooling and accounted for the radiation pressure from the stars, using 1D hydro-dynamical simulations. In this way, they predict a bubble size that is smaller by $\sim 30\%$ at an age of a few Myr and a temperature roughly one order of magnitude smaller than the one estimated in the adiabatic approximation. These changes, though sizeable, are not such as to change qualitatively the picture for CR transport with respect to the adiabatic case. As discussed in 
\cite{BlasiMorlino2023} this remains true even accounting for the presence of clumps, provided the number and density in the clumps are assumed be in a range that is considered reasonable. While we will comment further on the role of clumps below, in the remainder of this work we will adopt the assumption of adiabatic expansion of the bubble. 

Following \cite{Weaver+1977} and \cite{Gupta+2018}, \cite{Morlino+2021} provided some useful approximations for the position the forward shock (FS) and the termination shock (TS), that we use here. The position of the FS is at
\beq \label{eq:Rbubble}
 R_b(t) = 139  ~ \rho_{0,10}^{-1/5} \dot{M}_{-4}^{1/5}v_{8}^{2/5} t_{10}^{3/5} ~\rm pc,
\eeq
where $\rho_{0,10}$ is the ISM density in the region around the star cluster in units of 10 protons per cm$^{3}$, $v_{8}=v_{w}/(1000~\rm km\, s^{-1})$, $\dot M_{-4}=\dot M/(10^{-4}\rm M_{\odot}\,yr^{-1})$ and $t_{10}$ is the dynamical time in units of 10 million years. The wind luminosity is then $L_{w}=\frac{1}{2}\dot M v_{w}^{2}$. 
The termination shock is located at 
\beq \label{eq:RTS}
 R_{s}= 24.3 ~ \dot{M}_{-4}^{3/10} v_{8}^{1/10} \rho_{0,10}^{-3/10} t_{10}^{2/5} ~\rm pc \,.
\eeq

The accuracy of these expressions compared with more detailed calculations is at the level of $\lesssim 10\%$ \citep{Weaver+1977}.  We stress again that the speed of the TS in the laboratory frame is very low, so that the entire bubble structure evolves slowly and can be considered as stationary to first approximation. 

The wind speed and the mass loss rate can be estimated using momentum and mass conservation of the collective wind of all stars. We rely on the approach developed by \cite{Menchiari2023} who adopted the description of stellar wind velocity from \cite{Kudritzki-Puls:2000} and mass loss rate from \cite{Nieuwenhuijzen-deJager:1990}.

Given the assumption of adiabatic evolution of the bubble, the gas density downstream of the TS is approximately constant, hence, from mass conservation it follows that the plasma velocity drops as $1/r^2$. These scaling relations, previously described in detail by \cite{Morlino+2021}, are used below in solving the transport equation for non-thermal particles. 

The winds of the $\sim 100-1000$ young stars in the core are expected to interact strongly and dissipate some of the kinetic energy in the form of 
turbulent magnetic field. If a fraction $\eta_B$ of the kinetic energy of the wind is converted to magnetic energy, at the termination shock one may expect a turbulent magnetic field of order 
\beq \label{eq:B_TS}
B(R_{s})=7.4\times 10^{-6} \eta_{B}^{1/2} \dot M_{-4}^{1/5}v_{8}^{2/5} \rho_{10}^{3/10} t_{10}^{-2/5}~ G.
\eeq
This dissipation of kinetic energy into magnetic energy likely results in turbulence with a typical scale $L_c$ that is expected to be of order the size of the star cluster, $L_c\sim R_c\sim 1\div 2$ pc. 

Assuming that the turbulence follows a Kraichnan cascade, the diffusion coefficient upstream of the TS can be estimated as 
\begin{flalign}  \label{eq:DKra}
 D(E)\approx\frac{1}{3} r_{L}(p) v \left(\frac{r_{L}(p)}{L_{c}}\right)^{-1/2} = 
 1.1\times 10^{25} 
 \left( \frac{L_c}{1\rm pc}\right)^{1/2} \hspace{0.9cm} \nn \\
  \eta_{B}^{-1/4} \dot M_{-4}^{-1/10} v_{8}^{-1/5} \rho_{10}^{-3/20} t_{10}^{1/5} E_{\rm GeV}^{1/2}~\rm cm^2~s^{-1},
\end{flalign}
where $r_{L}(p)=pc/e B(r)$ is the Larmor radius of particles of momentum $p$ in the magnetic field $B(r)$. 

The effects of different turbulent cascade models have been presented in \cite{Morlino+2021}, mainly in terms of the effects on the spectrum of accelerated particles. A Kolmogorov-like turbulence is not considered here, because it leads to a particle spectrum that is difficult to reconcile with the observed gamma-ray spectrum, at least in the case of the Cygnus cluster \citep{menchiariPhD:2023}. On the other hand a Bohm-like diffusion is difficult to achieve: it can either result from CR self generated turbulence or from the superposition of injection at different scales with the same power. The former scenario has been discussed in detail in \cite{BlasiMorlino2023}, where the authors pointed out that it is unlikely for CR to produce self-generated perturbations in the environment described above. The latter scenario, instead, remains unexplored, even if it probably requires a significant level of fine tuning.

Downstream of the termination shock, we assume that the magnetic field is only compressed by the standard factor $\sqrt{(2R^2+1)/3}$, that for a strong shock (compression factor $R=4$) becomes $\sqrt{11}$. In this case $D_{2}\approx 0.55 D_{1}$. Clearly the downstream diffusion coefficient can be smaller than this estimate suggests, if other processes (such as the Richtmyer-Meshkov instability)   lead to enhanced turbulence behind the shock \cite[see, e.g.][]{Giacalone:2007p962}. We will discuss some implications of this scenario below. Finally we point out that in the downstream region the spatial dependence of the diffusion coefficient is very poorly constrained. Although here we assume for simplicity that the turbulence level stays constant in the whole downstream region, it is possible that damping effects may considerably increase the diffusion coefficient far away from the shock, especially in the absence of fresh injection of turbulence. We will explore the implications of this scenario in future work, since it requires a number of model dependent assumptions.

The functional form of the diffusion coefficient as in Equation~\eqref{eq:DKra} is expected to hold up to energies for which the Larmor radius equals the magnetic field coherence scale $L_c$. At larger energies the standard scaling, $D(E)\propto E^2$, arises, as can be found both analytically and using simulations of test particle transport in different types of synthetic turbulence  \cite[ee, e.g.][and references therein]{Subedi2017,Dundovic2020}.
In the calculations discussed below we will use the diffusion coefficient in Equation~\eqref{eq:DKra} (with a transition to $\propto E^2$ at high energies). 

Although, as discussed in Sec. \ref{sec:transport}, particle acceleration is supposed to occur mainly at the termination shock, it is important to keep in mind that in the downstream region particles can undergo additional acceleration, due to interactions with the same turbulence responsible for spatial diffusion, the so-called {\it second order} Fermi acceleration. This phenomenon was discussed in detail by \cite{Vieu+2022}, in a different scenario of particle energization in star clusters. The authors point out that effective second order acceleration also leads to depletion of the power in the form of high wavenumber turbulence, which in turn increases the diffusion coefficient of low energy particles, thereby  facilitating their escape from the downstream region. If this phenomenon is very efficient the grammage accumulated in the downstream becomes smaller, and both spallation and inelastic pp collisions become inefficient. It is worth noticing that while spatial scattering is not very sensitive to the isotropy of motion of perturbations, stochastic acceleration is only efficient when there is roughly the same number of waves moving in all directions. This makes the modeling of this process somewhat hard, especially outside the boundaries of a quasi-linear treatment and we decided here to exclude it from our calculations, though being aware of the potential implications of this assumption.

%%% SECTION %%%
\section{Acceleration and transport of CR protons and nuclei}
\label{sec:transport}

The transport of protons in the bubble excavated by the wind is described by the transport equation introduced and solved in \cite{BlasiMorlino2023}, which reads:
\bea
\frac{\partial}{\partial r}\left[ \tilde u r^2 N - r^2 D \frac{\partial N}{\partial r}\right] = E\frac{\partial N}{\partial E}\left[ \frac{1}{3}\frac{d(\tilde u r^2)}{dr} - r^2 \frac{b}{E}\right] + \nn \\
+ N \left[ \frac{1}{3}\frac{d(\tilde u r^2)}{dr} - r^2 b'\right],
\label{eq:transportE}
\eea
where $N(E,r)dE=4\pi p^2 f(r,p) dp$ is the distribution function of non-thermal particles as a function of energy, $f$ being the distribution function in phase space. For simplicity, below we focus on relativistic particles, so that $E\simeq pc$. In Equation~\eqref{eq:transportE} we omitted, for simplicity, the source term. The normalization is calculated {\it a posteriori} in terms of an efficiency of conversion of the ram pressure $\rho_w v_w^2$ into pressure of non-thermal particles. 
We introduced the rate of energy losses $b(E,r)=c \dot p$, and its derivative with respect to energy $b'(E,r)=\partial b(E,r)/\partial E$. For high energy protons energy losses are dominated by inelastic pp scattering \cite[see][]{BlasiMorlino2023}.
The effective velocity felt by particles, $\tilde u= u + \eta  v_A$, is the sum of the plasma speed $u(r)$ and the net speed of the waves responsible for the particle scattering expressed as $\eta$ times the Alfv\'en speed $v_A$. When the parameter $\eta$ is chosen to be zero (equal number of waves moving in both directions), the effect of scattering centers disappears. 

For nuclei, the equation is similar but the main channel responsible for particle evolution is due to spallation reactions, treated as catastrophic losses. Hence the transport equation for nuclei of type $\alpha$ can be written as:
\bea
\frac{\partial}{\partial r}\left[ \tilde u r^2 N_\alpha - r^2 D \frac{\partial N_\alpha}{\partial r}\right] = E\frac{\partial N_\alpha}{\partial E}\frac{1}{3}\frac{d(\tilde u r^2)}{dr} + \nn \\
+ \frac{1}{3}\frac{d(\tilde u r^2)}{dr}N_\alpha - \frac{N_\alpha}{\tau_{\rm sp,\alpha}}+\sum_{\alpha'>\alpha} \frac{N_{\alpha'}}{\tau_{\rm sp,\alpha'\to\alpha}}.
\label{eq:trNuclei}
\eea
Here $E$ denotes the energy per nucleon of the nucleus of type $\alpha$. Since we are still focusing on relativistic particles we safely neglected energy losses of the nuclei and included only spallation, with a characteristic time  $\tau_{\rm sp,\alpha}$. The time scale $\tau_{\rm sp,\alpha'\to \alpha}$ relates the spallation of a heavier nucleus to a nucleus of type $\alpha$. 

For a detailed prediction of the physical behaviour of nuclei in the environment of a star cluster, a full treatment of the nuclear cascade 
from any heavier nucleus to the nuclear specie $\alpha$ should be employed. However here we are only interested in making a physical point: the spectrum of nuclei is made harder by the effect of spallation. In order to make this point we focus on a few simpler cases: 1) we assume that $^4$He is injected at the shock (a reasonable assumption since $^3$He is secondary in nature) and that it fragments to $^3$He. We solve the transport equation, in its stationary form, for both species. We recall that the He spectrum that most experiments refer to is the total flux, to be interpreted as the sum of $^3$He and $^4$He. 2) We consider the cases of O and Fe nuclei that, to a good approximation, can be considered as pure primary nuclei and use this calculation to make a quantitative assessment of the role of spallation for nuclei heavier than He. 

In Equation~\eqref{eq:trNuclei}, the total cross section for the spallation of a nucleus of mass number $A$ has been taken as $\sigma_A=45 A^{0.7}$mb. Here $^3$He has been considered as purely secondary product of spallation reactions of $^4$He, although the branching ratio of this channel has been increased artificially to $0.75$ to mimic the contribution of heavier nuclei that have not been included explicitly in the spallation chain. In other words $\sigma({}^4{\rm He} \to {}^3{\rm He})= 0.75\sigma_4$.

We solve Equation~\eqref{eq:transportE} using a mixed technique, numerical and iterative, as introduced and discussed in \cite{BlasiMorlino2023} for the case of protons. For a given ansatz on the solution at the shock, $N_0(E)$\footnote{We initialize $N_0(E)$ using a pure power law with the expected slope, but we checked that the result does not change by changing this initial guess.}, Equation~\eqref{eq:transportE} is solved numerically upstream and downstream using a  grid discretized in radius and energy. The two solutions are matched at $r=R_s$ so that $N=N_0$. Such a solution is used to determine $D\frac{\partial N}{\partial r}|_{1,2}$ as functions of energy. At this point one can introduce 
\beq
D_{1,2} \frac{\partial N}{\partial r}|_{1,2} = \Dif_{1,2}(E) \tilde u_{1,2} N_0(E),
\eeq
so that Equation~\eqref{eq:transportE} becomes:
\beq
E\frac{dN_0}{dE} = N_0 \left[ -\frac{3\Dif_2(E)}{1-\erre} + \frac{3\erre}{1-\erre}(\Dif_1(E)-1) - \frac{\erre+2}{\erre-1}\right],
\label{eq:shock}
\eeq
where we introduced the compression factor of velocities of the scattering centers at the TS, $\erre=\tilde u_1/\tilde u_2$. The solution of Equation~\eqref{eq:shock} can now be written as
\beq
N_0(E)=K E^{-\frac{\erre+2}{\erre-1}} \exp\left\{\int_0^E\frac{dE'}{E'} 
\left[-\frac{3\Dif_2}{1-\erre} + \frac{3\erre}{1-\erre}(\Dif_1-1)\right]\right\}.
\label{eq:N0}
\eeq
It is useful to notice that in the case of a plane parallel shock, $\Dif_1\to 1$ and $\Dif_2\to 0$, so that the solution reduces to the standard power law $N_0\propto E^{-\frac{\erre+2}{\erre-1}}$. The exponential term in Equation~\eqref{eq:N0} takes into account both the spherical symmetry and the effect of energy losses through $\Dif_1$ and $\Dif_2$. The specific energy dependence of $\Dif_{1,2}$ shapes the spectrum of accelerated particles as a result of proximity to the maximum energy and because of the spherical topology of the outflow. 

Equation~\eqref{eq:trNuclei} is solved in a similar way for primary nuclei. For secondary nuclei the situation is slightly more complicated: in such a case there is no injection term at the shock, while the production term is spread in space due to spallation. Because of this, secondary nuclei produced in the region upstream of the termination shock are advected toward the shock and are subject to some level of acceleration, an effect similar to that described for SNRs by \cite{Blasi2009} for electron-positron secondary pairs, by \cite{MertschSarkar2009,Berezhko2003} for secondary boron and by \cite{BlasiSerpico2009} for antiprotons. 
In order to correctly describe this effect it is useful to relate the derivative upstream of the shock to the production rate upstream. To do so, we integrate the transport equation from $r=0$ to $r=R_s$ to obtain:
\begin{equation}
    \left.D\frac{\partial N_\alpha}{\partial r}\right|_1 = \tilde u_1 N_{\alpha,0} + \delta u N_{\alpha,0} - R_s \int_0^1 d\xi \frac{\xi^2 N_\alpha}{\tau_{\rm sp,\alpha}},
\end{equation}
where $\xi \equiv r/R_s$ and we introduced the quantity:
\begin{eqnarray}
\delta u = -\frac{2}{3}\frac{\tilde u_1}{N_{\alpha,0}} \int_0^1 d\xi N_\alpha \xi \left(1-q_\alpha(E,\xi)\right) + \nn \\
+ \frac{R_s}{N_{\alpha,0}} \int_0^1 d\xi \frac{\xi^2 N_{\alpha'}}{\tau_{\rm sp,\alpha'\to \alpha}},
\end{eqnarray}
and $q_\alpha(r,E) \equiv -d\ln N_\alpha/d\ln E$ is the slope of the function $N_\alpha$ at location $r$ and energy $E$.

Integrating the transport equation around the shock location we obtain:
\begin{equation}
    E\frac{dN_{\alpha,0}}{dE} = -s \, N_{\alpha,0} + \frac{3N_{\alpha,0}}{r-1}\left[ {\cal D}_2 - \frac{\delta u}{\tilde u_2}\right] + F(E),
\label{eq:N0Nuclei}
\end{equation}
where 
\begin{equation}
    F(E) = \frac{3R_s}{\tilde u_2 (r-1)}\int_0^1 d\xi \frac{\xi^2 N_{\alpha'}}{\tau_{\rm sp,\alpha'\to\alpha}}.
\end{equation}
Here we used explicitly the compression factor $r=\tilde u_1/\tilde u_2$ and the slope $s=(r+2)/(r-1)$ that would be obtained for a plain parallel shock. The function $F(E)$ plays the role of a source function for nuclei of type $\alpha$. The formal solution of Equation~\eqref{eq:N0Nuclei} can be written in an implicit way as follows:
\begin{eqnarray}
    N_{\alpha,0}(E) = \int_0^E \frac{dE'}{E'} F(E') \left( \frac{E}{E'}\right)^{-s} \times \nn \\
    \exp \left\{ \int_{E'}^E \frac{dE"}{d E"} \left[ \frac{3}{r-1}\left({\cal D}_2-\frac{\delta u}{\tilde u_2}\right)\right]\right\}.
\end{eqnarray}
The solution as written here is formally reminiscent of that found in problems of reacceleration. In fact, the energization of secondary nuclei produced upstream can be considered as a sort of reacceleration. The procedure for solving the overall problem of transport of secondary nuclei is basically identical to the one discussed above for primary nuclei. For a given ansatz on $N_{\alpha,0}$ one can solve for the spatial transport numerically and so evaluate from the calculation the quantities ${\cal D}_2$, $\delta u$ and $F(E)$. This allows us to evaluate an updated $N_{\alpha,0}$. A simple iterative technique leads to the solution. 

\subsection{Limitations of a spherically symmetric approach}
\label{sec:limits}

It may be argued that a spherically symmetric approach adopted here may lose track of the fact that the actual target for CR interactions is in the form of dense clouds that fill a limited fraction of the volume of the bubble. In order to make a quantitative assessment of this effect we estimate here the fraction of cosmic rays that in a star cluster may be subject to the grammage contributed by the clouds. 

Since the wind density is typically very low, the average density $n$ is mainly contributed by the clouds and we can write:
\begin{equation}
    \frac{4\pi }{3} R_b^3 \, n \, m_p = N_{\rm cl} M_{\rm cl} = N_{\rm cl} \frac{4 \pi}{3} R_{\rm cl}^3 \, n_{\rm cl} \, m_p ,
\end{equation}
where $n_{\rm cl}$ and $R_{\rm cl}$ are the typical density and radius of a cloud of mass $M_{\rm cl}$. If follows immediately that the number of clouds in the bubble is:
\begin{equation}
    N_{\rm cl}=\frac{n}{n_{\rm cl}}\left( \frac{R_b}{R_{\rm cl}}\right)^3.
    \label{eq:Ncl}
\end{equation}
It is worth noticing that for fiducial values of the parameters, $n=10\, \rm cm^{-3}$, $n_{\rm cl}=500 \, \rm cm^{-3}$, $R_b=100\, \rm pc$ and $R_{\rm cl}=2\, \rm pc$, one may expect $N_{\rm cl}\approx 2500$, which implies a typical distance between two clumps of $d\sim 4 R_{\rm cl}$. It is instructive to estimate the time necessary to travel through diffusion between two clumps, $\tau_c$, compared with the two relevant time scales, the advection and diffusive time scale:
\begin{equation}
    \tau_c=\frac{d^2}{6D(E)}=16 \frac{R_{\rm cl}^2}{6D(E)}=16 \left( \frac{R_{\rm cl}}{R_b}\right)^2
    \frac{\tau_{\rm dif}(E)}{\tau_{\rm adv}}\tau_{\rm adv},
    \end{equation}
where $\tau_{\rm adv}$ and $\tau_{\rm dif}$ are respectively the advection and diffusion time scales in the bubble. These quantities can be read off Figure~\ref{fig:Times}, together with other relevant time scales discussed in the next section, for a star cluster with $\dot M=1.5\times 10^{-4}~\rm M_\odot\,  yr^{-1}$, $v_w=2800\, \rm km\,s^{-1}$, age of 3 Myr, a mean density in the bubble $n=10~\rm cm^{-3}$, external density of 20 $\rm cm^{-3}$ and $\eta_B=0.1$. It may be easily seen that $\tau_c$ is much shorter than both the advection and diffusion times. This implies that whether the particles accelerated at the termination shock escape the bubble advectively (low energies) or diffusively (high energies), they are bound to cross many dense clumps, and the accumulated grammage really depends upon the time spent inside the clumps rather than in the low density part of the bubble. For instance, for a particle that escapes from the cavity through advection, the transverse diffusive motion in one advection time is 
\begin{equation}
    \sim \sqrt{D \tau_{\rm adv}}=R_b \sqrt{\frac{D \tau_{\rm adv}}{R_b^2}} = R_b \sqrt{\frac{\tau_{\rm adv}}{\tau_{\rm dif}}},
\end{equation}
which is a large fraction of the radius of the bubble. In fact the volume probed by the particles while advecting toward the edge of the bubble 
is $\sim \pi R_b^3 (\tau_{\rm adv}/\tau_{\rm dif})$, which even at energies as low as 10 GeV is of order $\sim 2\%$ of the entire volume, meaning that about 50 clumps are encountered before escape. When the particles escape diffusively, by definition they probe the entire volume of the bubble and cross all the clumps in it. 

Let $\tau_{\rm esc}$ be the escape time, defined as the minimum between the advection and diffusion time scales. The grammage that the particles traverse can then be written as 
\begin{equation}
    X(E)=n_{\rm cl} c \, m_p \, \tau_{\rm esc} \,f,
\end{equation}
where $f$ is the fraction of time spent in the clumps. Here, we assume this fraction to be equal to the volume filling factor of clumps\footnote{ This assumption can be shown to be rigorously valid for a 1D geometry under the condition that $D_{\rm cl} \gg D_{\rm b} R_{\rm cl}/R_{\rm b}$, where $D_{\rm b}$ is the diffusion coefficient in the bubble.}:
\begin{equation}
    f= \frac{N_{\rm cl}V_{\rm cl} \frac{V}{V_b}}{V}
    = N_{\rm cl}\left(\frac{R_{\rm cl}}{R_b}\right)^3 
    = \frac{n}{n_{\rm cl}},
    \label{eq:f}
\end{equation}
where $V$ is the volume accessible to particles during the escape time (when escape is dominated by advection, $V<V_b$), $V_b=\frac{4}{3}\pi R_b^3$ is the total bubble volume and
$V_{\rm cl}=\frac{4}{3}\pi R_c^3$ is the volume of an individual clump. In the last part of Equation~\eqref{eq:f} we used Equation~\eqref{eq:Ncl}. Replacing this expression in the grammage we find that 
\begin{equation}
    X(E) = n \, c \, m_p \, \tau_{\rm esc},
\end{equation}
which shows that the introduction of the mean density in the bubble correctly describes the fact that there are concentrated clumps of dense material in the same region. This could be expected since, as discussed above, at all energies the number of clumps traversed by the particles is $\gg 1$, both in the advection dominated and diffusion dominated regimes. We conclude that the calculations discussed here, adopting a spherically symmetric geometry and a mean density inside the bubble, properly describe the effect of spallation of nuclei of the spectrum of particles, as long as $N_{\rm cl}\gg 1$. 

If only a few dense clouds of cold gas are present inside the bubble, it is possible that some of the particles escape with no interactions with clouds, while the ones that happen to cross a clump may suffer severe spallation. This situation is not contemplated here.

The estimates illustrated above are physically justified when particles spend a short time inside individual clumps compared with the total escape time from downstream, which requires the transport in individual clumps to correspond to a diffusion coefficient equal to or larger than the one outside the clumps. In particular the estimate is justified in the case of free streaming inside clumps, which may be the case when effective damping of small scale turbulence inside clumps leads to a suppression of scattering.

%%% SECTION %%%
\section{Results}
\label{sec:results}
Here we illustrate our results for the cases introduced above. We stress that the purpose of this section is only to show the physical effect of accounting for nuclear spallation in the bubble, while a more quantitative analysis would require to introduce the whole chain of spallation reactions from heavy to lighter nuclei and to consider source terms for all primary nuclei, depending on the efficiency of particle injection at the termination shock. These efficiencies would require to be fitted to the observations at the Earth, after propagation of the CR spectra escaping star clusters on Galactic scales.

\begin{figure}
\centering
\includegraphics[width=.45\textwidth]{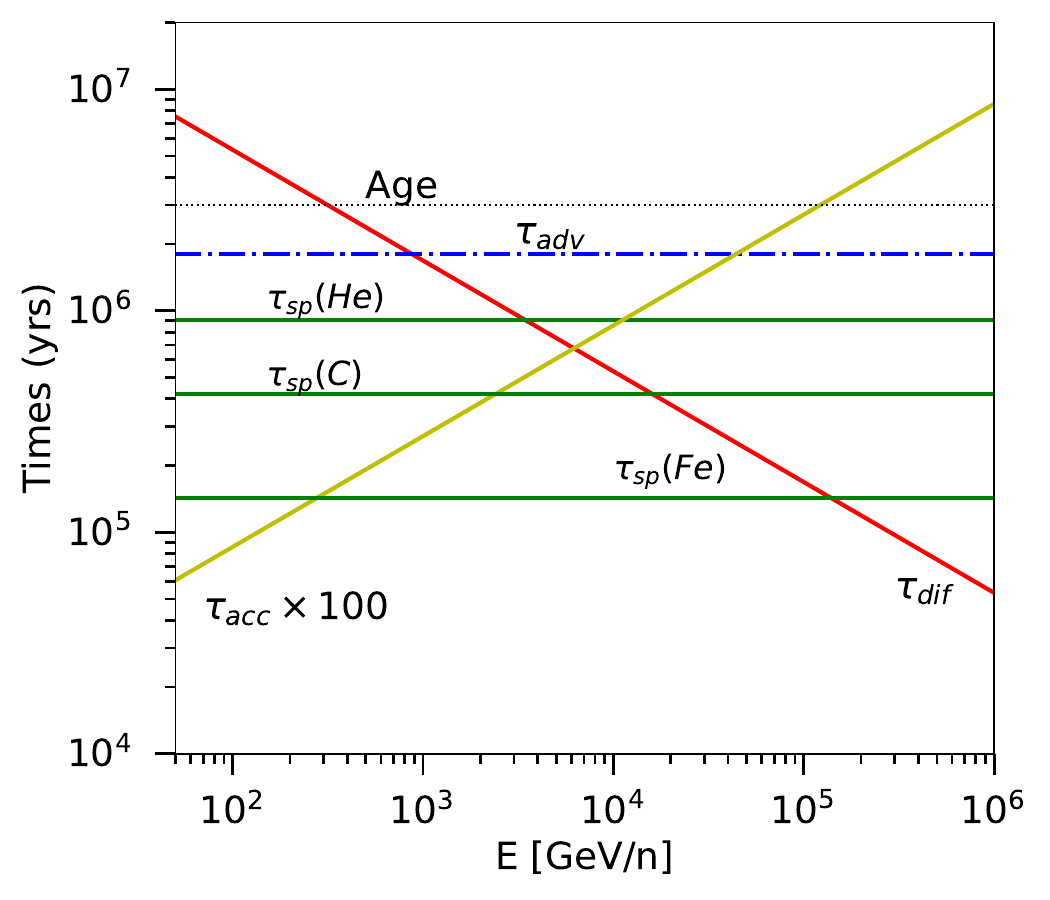}
\caption{Time scales for diffusive escape from the bubble ($\tau_{dif}$), advection ($\tau_{adv}$), acceleration ($\tau_{acc}$), multiplied here by 100 to make it visible in the same plot, and spallation for He, C and Fe nuclei. A reference value of $n=10~\rm cm^{-3}$ for the mean gas density in the bubble has been adopted. These timescales are compared with the age of the star cluster (dotted black line). We assumed $\dot M=1.5\times 10^{-4}~\rm M_\odot/yr$, $v_w=2800$ km~s$^{-1}$, external density of 20 $\rm cm^{-3}$ and $\eta_B=0.1$.}
\label{fig:Times}
\end{figure}

The time scale for spallation of He, C and Fe nuclei in a typical bubble resulting from the interaction of the star cluster with the surrounding ISM is shown in Figure~\ref{fig:Times}, together with the diffusion, advection and acceleration time scales (the latter only for protons) and the age of the cluster, assumed here to be $3$ million years. The gas density in the bubble adopted here as a benchmark value, as due to dense clouds, is $n=10~\rm cm^{-3}$. Figure~\ref{fig:Times} illustrates the situation very clearly: first, particle acceleration is very fast compared to the typical dynamic scale of the cluster, as one can see by comparing advection and diffusion time scales with the acceleration time, estimated here as
\begin{eqnarray}
    \tau_{\rm acc}(E)=\frac{3}{u_1-u_2}\left[ \frac{D_1(E)}{u_1}+\frac{D_2(E)}{u_2}\right].
\end{eqnarray}
On the other hand the equality of the diffusion length downstream to the size of the termination shock is the criterion that defines the maximum energy reached by the particles \cite[]{Morlino+2021}.

Spallation reactions occur faster than advection for elements heavier than He. For He, spallation and advection are competing processes, although diffusion eventually takes over spallation reactions for energies larger than $\sim 5$ TeV/n. For elements heavier than He the role of spallation becomes correspondingly more important and represents the main factor in the evolution of the density of these CR elements in the downstream region. 

These processes all concur to shape the spectrum of CR species in the bubble and eventually the spectrum of particles escaping the bubble itself, a quantity that is especially important since it plays the role of injection term of nuclei in the description of CR transport in the Galaxy. Below we illustrate our results for He nuclei, with special emphasis on the role of $^3$He, and some heavier nuclei (O and Fe). 

%%% SUBSECTION %%%
\subsection{The case of helium nuclei}
\label{sec:helium}
While measurements of the spectrum of helium nuclei at high energies refer to the sum of fluxes of $^3$He and $^4$He, from the point of view of the physical origin of these elements, the former is basically all secondary product of spallation, while the latter is mainly a primary nucleus, weakly affected by the spallation of heavier elements. In the following we shall make the simple assumption that the spallation of $^4$He mainly leads to production of $^3$He and that the contribution to $^3$He from heavier elements can be mimicked by enhancing by hand the cross section of production from $^4$He by a factor $\sim 2.5$ \cite[]{Coste2012} (see also Section \ref{sec:transport}). This assumption has its limitations since we know that in the Galaxy part of $^3$He comes from the fragmentation of C, O and heavier elements. However, as will be clear in \S~\ref{sec:heavy}, spallation in the environment of a star cluster destroys most of the heavy CR components, hence these limitations are less severe than they would be in other contexts, such as the propagation in the Galaxy. On the other hand, it is possible that severe spallation on heavy elements may contribute to the amount of He (as well as, in principle, protons). 

With these assumptions, we can estimate the spectra of $^3$He and $^4$He produced in a star cluster and the spectra at escape, to be considered as effective source spectra. We also compare the escape spectrum of $^3$He+$^4$He with that of protons.

Upstream of the termination shock, $^4$He nuclei accelerated at the shock can diffuse a distance of order $D_1/\tilde u_1$ away, and this is also the region where the production of secondary $^3$He nuclei can take place. Since high energy particles can diffuse further away from the shock, more high energy $^3$He is also produced, namely the injected spectrum of secondary $^3$He nuclei that reach the shock from upstream is harder than the spectrum of the parent $^4$He nuclei. The $^3$He nuclei produced upstream are automatically injected as non-thermal particles and get energized through diffusive shock acceleration, an effect reminiscent of what is expected for secondary electron-positron pairs at SNR shocks \cite[]{Blasi2009}. 

On the other hand, downstream of the shock the production of secondary nuclei can take place anywhere before reaching the free escape boundary. It is then clear that most $^3$He is produced downstream. In turn, $^3$He also undergoes spallation reactions downstream, which deplete its flux, especially at low energies where transport is dominated by advection. At the highest energies the escape from dowsntream is dominated by diffusion and these particles suffer less fragmentation. 

\begin{figure*}
\centering
\includegraphics[width=1.0\textwidth]{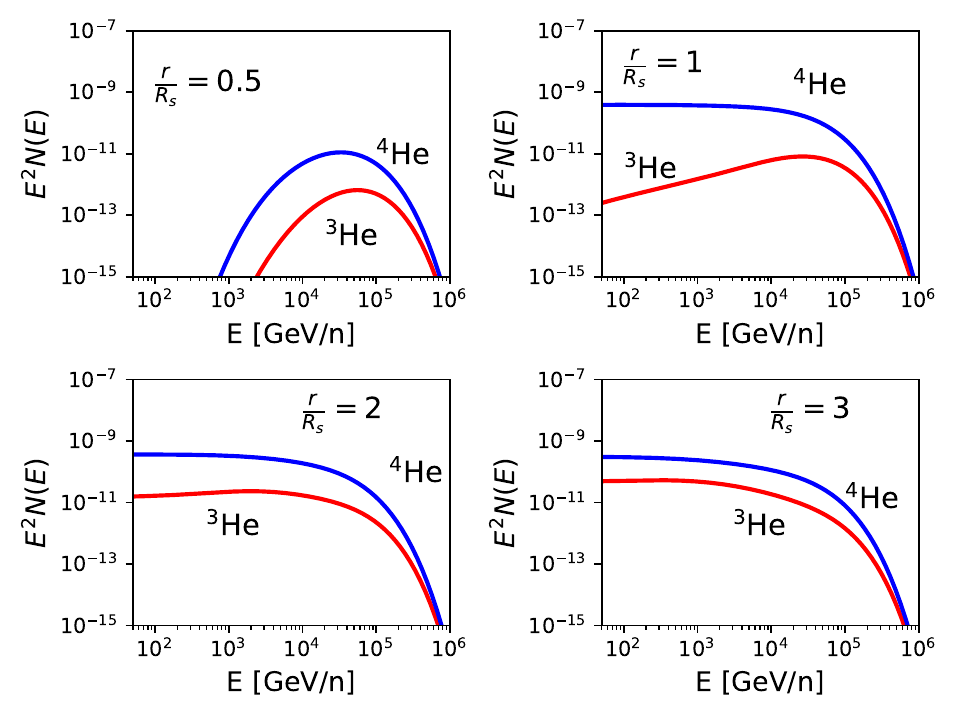}
\caption{Spectrum of $^4$He and $^3$He (as $E^2 N(E)$ in arbitrary units here) at different locations. From top left and clockwise, the panels show the cases $r/R_s=0.5$, $r/R_s=1$ (shock location), $r/R_s=2$ and $r/R_s=3$.}
\label{fig:SpecSpace}
\end{figure*}

The situation is illustrated in Figure~\ref{fig:SpecSpace}, where different panels show the spectrum of $^4$He and $^3$He at different locations from $r/R_s=0.5$ (upstream) to $r/R_s=3$ (downstream). The spectrum of $^4$He accelerated at the shock is shown in the panel labelled as $r/R_s=1$. We adopted typical parameters of a young massive stellar cluster: $v_w=2800$ km/s and $\dot M=1.5\times 10^{-4}~\rm M_\odot/yr$, compatible with the values found by \cite{Menchiari2023}. In addition we assumed a Kraichnan spectrum of the turbulence with coherence scale $L_c=1$ pc. With such a choice, the spectrum of $^4$He starts falling at an energy of $\sim 100$ TeV/n, an effect that we discussed in detail in \cite{Morlino+2021} and \cite{BlasiMorlino2023}. At lower energies the spectrum has the standard shape $\propto E^{-2}$, since we adopted $\eta=0$ in the definition of $\tilde u$. At the location of the termination shock, the only $^3$He present in the stationary situation is represented by nuclei produced upstream as a result of spallation of $^4$He and energized at the shock crossing. One should notice the peculiar extremely hard spectrum of $^3$He, typical of reacceleration of secondary particles (at least when the diffusion coefficient increases with energy, which is typically the case). 

For $r/R_s=0.5$ (upstream of the shock), only $^4$He nuclei with a diffusion length comparable with or larger than $\sim 0.5R_s$ can reach such a location, hence the spectrum of $^4$He is peaked at high energies. Since spallation conserves the energy per nucleon, the spectrum of $^3$He nuclei produced as spallation products is also peaked at similar values of the energy/nucleon. 

In the region downstream of the termination shock, two phenomena take place. On one hand $^4$He produces more $^3$He while advecting and diffusing further down toward the external part of the bubble. On the other hand $^3$He nuclei produced downstream are also transported through advection and diffusion, and suffer additional spallation reactions. In fact, one can see that the normalization of the spectrum of $^4$He slightly decreases while moving toward the edge of the bubble and the normalization of the spectrum of $^3$He correspondingly increases. 

Moreover while nuclei approach the outer boundary, it becomes more likely for them to escape the bubble diffusively. This combination of processes shapes the spectrum of particles escaping the remnant, determined as $-D\frac{\partial N}{\partial r}|_{r=R_b}$, as well as the spectra of nuclei at each location downstream. From the panels of Figure~\ref{fig:SpecSpace} referring to $r/R_s=2$ and 3, one can see that most of the production of $^3$He, especially at low energies, occurs in the downstream region. Eventually, the highest energy part disappears because these particles are the ones that most easily escape the bubble in a diffusive way. 

\begin{figure}
\centering
\includegraphics[width=.5\textwidth]{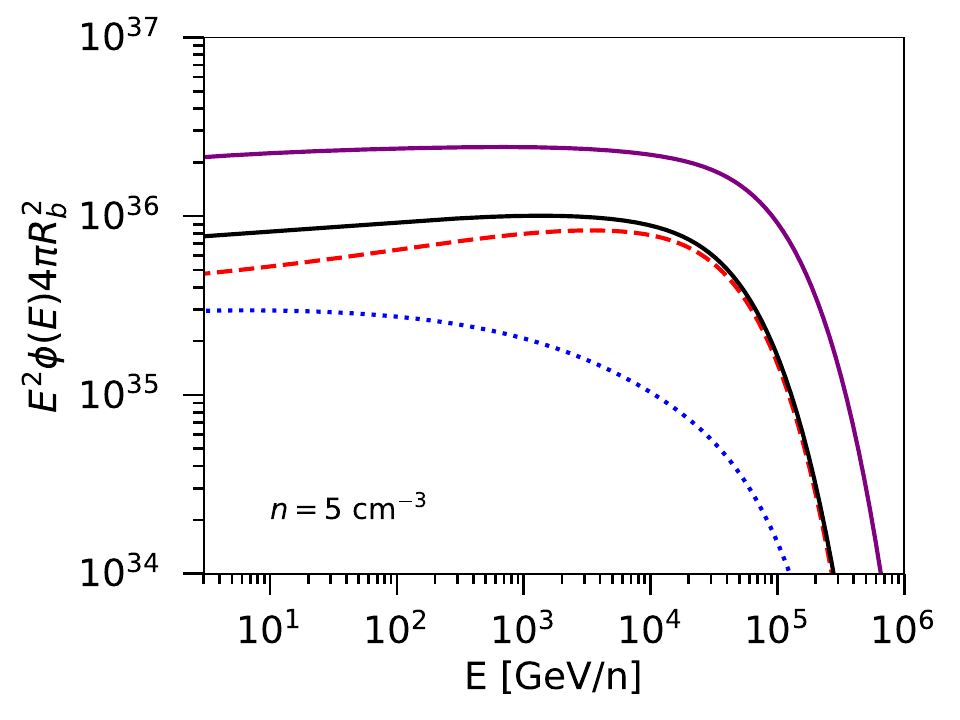}
\includegraphics[width=.5\textwidth]{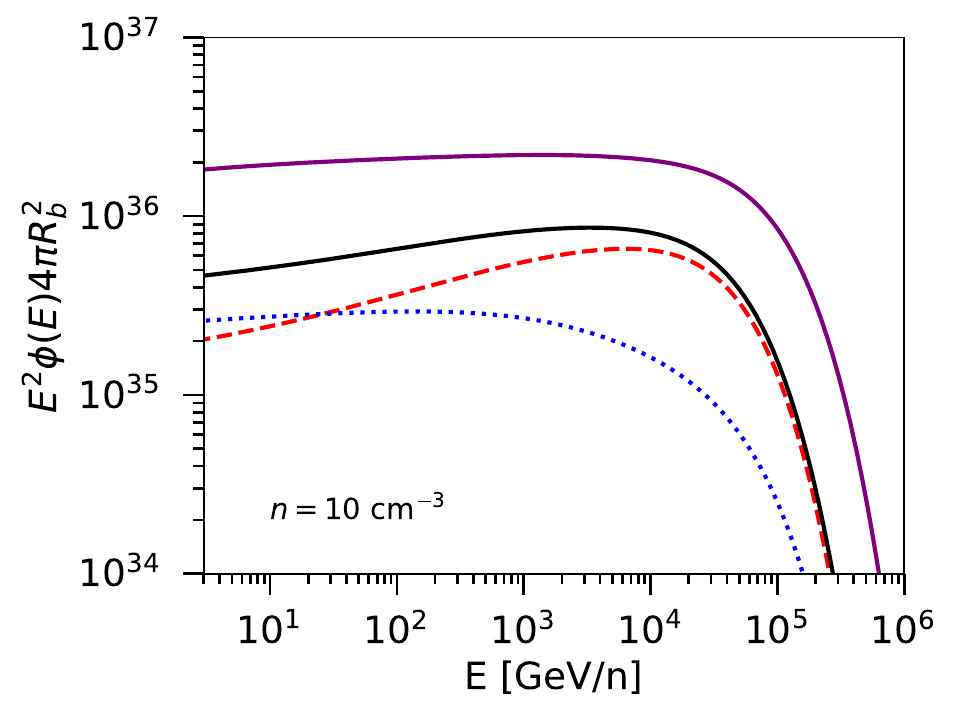}
\includegraphics[width=.5\textwidth]{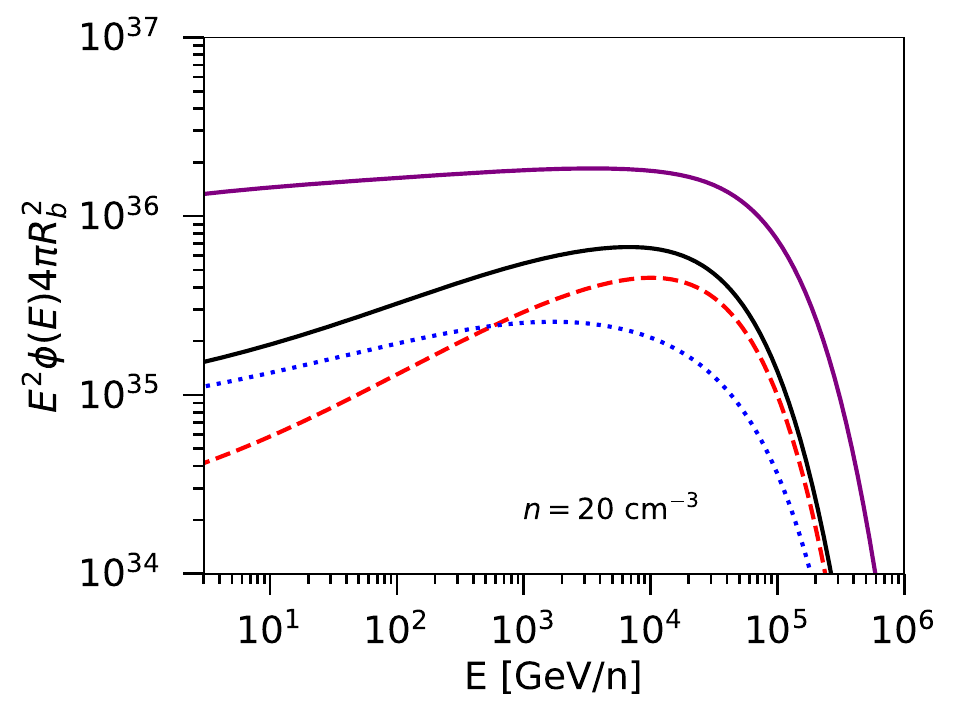}
\caption{Spectrum of escaping protons (purple solid line) compared with the escaping spectra of $^4$He (red dashed line) and $^3$He (dotted blue line) nuclei. The solid black line is the total He flux escaping the bubble. The top, middle and bottom panels refer respectively to a mean density $n=5~\rm cm^{-3}$, $n=10~\rm cm^{-3}$ and $n=20~\rm cm^{-3}$.}
\label{fig:Escape}
\end{figure}

The spectrum of $^4$He and $^3$He nuclei that leave the bubble is shown in Figure~\ref{fig:Escape} and compared with the spectrum of escaping protons, for three values of the mean gas density in the bubble, $n=5~\rm cm^{-3}$, $n=10~\rm cm^{-3}$ and $n=20~\rm cm^{-3}$. Since the experiments we have currently at our disposal can only measure the total flux of He, especially at high energies, we also show the total He flux. We stress that the relative normalization of the proton and He spectra is arbitrary, in that it is determined by the efficiency of acceleration of the different species and by their relative abundances. On the other hand, the relative normalization of the spectra of $^4$He and $^3$He nuclei is only fixed by the spallation cross sections and by the processes responsible for their transport, namely advection and diffusion. Hence such ratio, for given parameters of the problem, can be predicted. 

The comparison between the solid purple and black lines gives an immediate feeling of the effect of hardening of the spectrum of nuclei compared with that of protons escaping the bubble into the ISM. In the most reasonable case of density $n=10~\rm cm^{-3}$, the effect of losses is to harden the total He spectrum with respect to that of protons by $\sim 0.07$, that compares rather well with the effect required to explain the {\it discrepant hardening} in observations. For larger gas density, the effect becomes correspondingly larger and quickly becomes rather severe, as can be seen in the lower panel of Fig. \ref{fig:Escape}, with density $n=20~\rm cm^{-3}$. In the same way, a lower mean density in the bubble (top panel) leads to a less pronounced hardening of the He spectrum. Clearly all the considerations above apply here to the energy range where the spectra are reasonably close to power laws. As discussed here and in \cite{BlasiMorlino2023}, for typical values of parameters the effective maximum energy of accelerated particles is $\lesssim 100$ TeV, hence deviations from the power law trend are already visible in the tens of TeV energy range. This is also the range where the phenomenon of discrepant hardening has been unambiguously measured. 

\begin{figure}
\centering
\includegraphics[width=.5\textwidth]{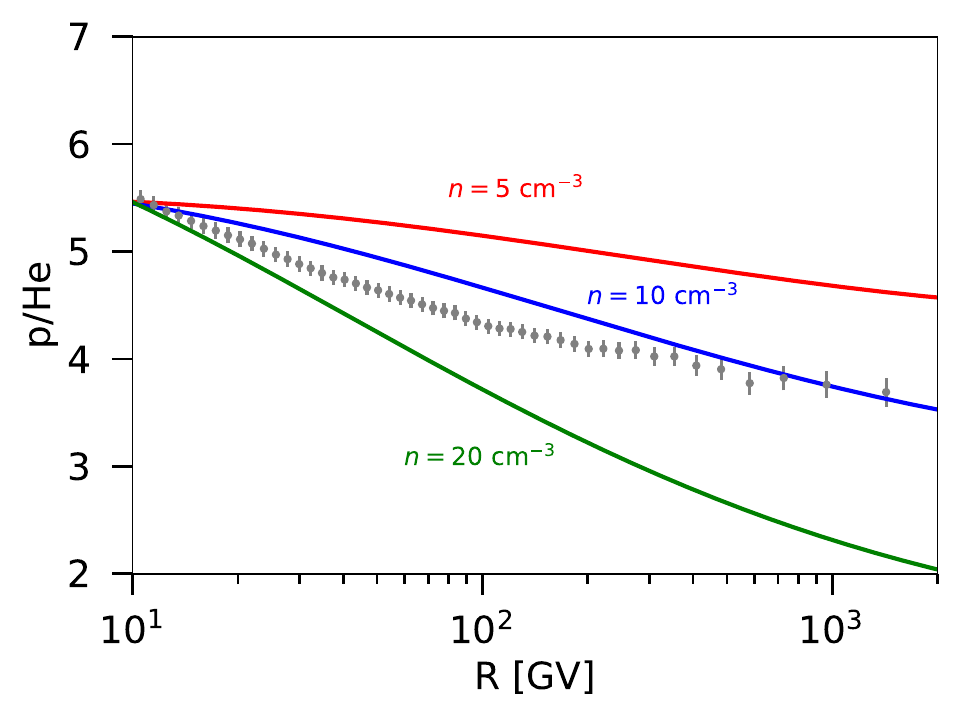}
\caption{Ratio of the spectra of protons and helium nuclei ($^4$He+$^3$He) escaping a star cluster for different values of the mean gas density in the bubble, $n=5~\rm cm^{-3}$ (red), $n=10~\rm cm^{-3}$ (blu) and $n=20~\rm cm^{-3}$ (green). The ratio has been normalized to be the same in the different cases at rigidity $10$ GV. Data from AMS-02 \citep{AMS-pHeRatio} are also shown in the same plot for the sole purpose of allowing a qualitative comparison.}
\label{fig:Ratio}
\end{figure}

Although, as discussed above, the relative ratio of fluxes of protons and helium is arbitrary here, it is instructive to show the energy dependence of the ratio $\rm p/He$ when the ratio is normalized to be the same at rigidity $10\, \rm GV$. This quantity is shown in Figure \ref{fig:Ratio} for the same values of the mean gas density adopted earlier. In Figure \ref{fig:Ratio} we also show the data of AMS-02 for the ratio \cite[]{AMS-pHeRatio} for the sole purpose of allowing the reader to achieve a qualitative comparison. One may notice that the ratio as computed here is in a qualitative good agreement with the observed trend, at least for mean density $\sim 10~ \rm cm^{-3}$. However, it is important to keep in mind that the data refer to the propagated spectra while we only calculated the ratio of fluxes escaping an individual star cluster. Although propagation does not affect dramatically the ratio for rigidities $\gtrsim 10 ~\rm GV$, it remains true that the physical meaning of the two ratios is formally different. Moreover, the ratio shown here, as well as the fluxes of nuclear species, refer to an individual star cluster, with parameter values resembling those of the Cygnus region, while the fluxes and the ratio should be computed using a proper superposition of clusters weighed with a luminosity function, a much more ambitious goal than the one aimed at here. Finally, the spectra of p and He computed here are to be interpreted only as indicative since we are not properly accounting for the production of lighter elements from the spallation of heavier elements. 

Despite all these caveats, we believe that these results prove the potentially strong effect of the star clusters environment on the spectra of the escaping protons and helium nuclei and may stimulate further investigation of the problem.

%%%SUBSECTION %%%
\subsection{The case of heavier nuclei}
\label{sec:heavy}

If the physical reason for the discrepant hardening between protons and He nuclei is the occurrence of spallation reactions in the bubble excavated by the collective stellar wind of the cluster, one should wonder what happens to heavier nuclei, characterized by larger spallation cross sections. The corresponding time scales for spallation become appreciably smaller than the advection time scale and the age of the cluster, so that the escape of heavier nuclei from the bubble excavated by the star cluster can easily become an issue. 

\begin{figure}
\centering
\includegraphics[width=.45\textwidth]{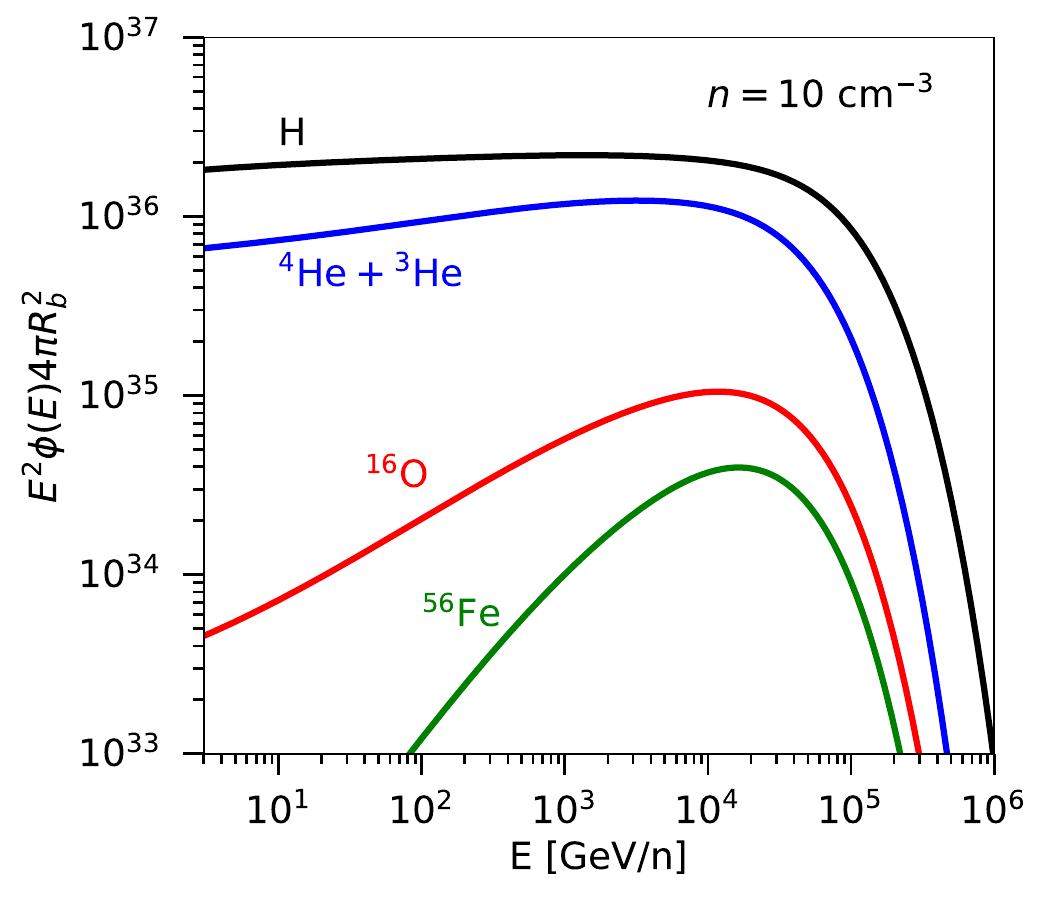}
\caption{Spectra of protons (black line), $\rm ^4He+^3He$ nuclei (blue line), oxygen (red line) and Fe (green line) escaping the bubble in the case of mean density inside the cavity $n=10~\rm cm^{-3}$.}
\label{fig:SpectraNuclei}
\end{figure}

As one can read off Figure~\ref{fig:Times}, the time scale for spallation of heavier elements (C, O, Fe) is so short that they are likely to be destroyed completely, even for relatively low mean gas densities, $n\lesssim 10~\rm cm^{-3}$. This is illustrated more clearly in Figure~\ref{fig:SpectraNuclei}, where we show the spectrum of protons, $^4$He+$^3$He nuclei, oxygen and Fe escaping the bubble in the case of mean density $n=10~\rm cm^{-3}$. One can appreciate again that while the spectrum of He is only slightly hardened, the effect on heavier nuclei is dramatic: for instance the spectrum of oxygen nuclei would be hardened by $\sim 0.5$ with respect to H, which in practical terms would mean that the oxygen flux at energies $\lesssim 10^4$ GeV would be severely depleted. An even more severe effect can be seen in the spectrum of Fe nuclei, for which only very high energy particles partially survive because they can escape diffusively. In fact, Figure~\ref{fig:Times} clearly shows that the spallation time for Fe nuclei is shorter than the diffusion time at all energies, though the two time scales become comparable within a factor of two, at energies $\gtrsim 10^6$ GeV/n, where however the spectra have already fallen down because of the maximum energy at the acceleration site. 

We stress once more that the physical information to be extracted from Figures~\ref{fig:Escape} and \ref{fig:SpectraNuclei} is in the shape of the spectra, not in their relative normalizations. The latter depend on the efficiency of acceleration of each species and their relative abundances and are clearly free parameters in our approach. If star clusters contribute to the flux of CRs in the Galaxy, then it appears clear that their main contribution is in the form of light elements, while heavier elements should be accelerated in more conventional environments, such as SNR shocks. On the other hand, since spallation seems to be so important, especially downstream of the termination shock, it seems urgent to carry out a calculation in which the full chain of spallation reactions, including all elements, is followed, so as to assess the contribution of such reactions to the flux of lighter elements and to the flux of intermediate secondary nuclei, such as B, Be and Li, that are produced and destroyed effectively in the bubble. 

%{\bf There are a few caveats that  to be applied to the conclusions illustrated above: the calculations carried out here, as well as in \cite{Morlino+2021} and \cite{BlasiMorlino2023}, are all in 1D spherical coordinates, namely only dependent  upon the radial coordinate. 

Here it is important to remind the reader once more that the mean density of gas adopted in the present calculation is to be interpreted as due to the presence of dense clouds inside the cavity. As discussed in \S\ref{sec:limits}, for typical density of gas in a cloud $n_{\rm cl}\sim$few hundred particles per $\rm cm^3$, the probability of CRs encountering a cloud is of order unity and the calculations above provide a proper description of the transport of CRs in the bubble. The calculations eventually fail only in the unlikely situation in which the gas is concentrated in only a handful of very dense clouds in the bubble. In this case, the CRs that happen to cross a cloud would suffer severe spallation, while the ones that do not cross a cloud would escape the SC unaffected by spallation. For typical values of the parameters this is not the case.

%%% SECTION %%%
\section{Conclusions}
\label{sec:concl}

While it has become clear that the bulk of the high energy gamma radiation observed from selected clusters, for instance the Cygnus OB2 association, is of hadronic origin, the role of star clusters as sources of high energy cosmic rays remains matter of much debate: it is often believed that these sources may accelerate particles to the PeV energy region more easily than SNRs. This may be true, in that the termination shock behaves as an efficient accelerator even in the absence of strong CR induced instabilities \cite[]{BlasiMorlino2023}. Moreover the nominal maximum energy can be as high as several PeV, but as discussed in \cite{Morlino+2021} and \cite{BlasiMorlino2023}, the actual spectrum of accelerated particles seems to become steep at energies much smaller than this nominal $E_{\rm max}$. This has to do with the energy dependence of the diffusion coefficient, which is probably the least known aspect of the problem. On the other hand, the case of Bohm diffusion, required to avoid or alleviate these problems, is expected only if CRs with a spectrum $\sim E^{-2}$ drive powerful streaming instability, a situation that is hard to conceive in star clusters \cite[see discussion in][]{BlasiMorlino2023}, or if turbulence in the wind region happens to be injected with roughly equal power on different scales over a broad range of scales. To be conservative, we make the statement that the role of star clusters (or at least of the bulk of star clusters) as PeVatrons remains to be assessed.

On the other hand, the gamma ray emission certainly shows that these are powerful CR accelerators. The gas column density measured in the direction of SCs in some specific cases, like Cyg OB2, also suggests that the density of local gas in the bubble is $\sim 10\rm \, cm^{-3}$ \citep{Aharonian+2019NatAs, Menchiari2023}. In this framework it is reasonable to wonder about the fate of protons and nuclei accelerated at the termination shock. This was the focus of the present article, where particle acceleration of protons and nuclei, with energy losses and fragmentation taken into account, was investigated in detail. 

The spectrum of protons was found to be affected by inelastic energy losses \cite[]{BlasiMorlino2023}, though to a modest extent for a gas density $\sim 10~\rm cm^{-3}$. This is consistent with the estimates of the advection and diffusion time scales compared with the time scale of losses. The situation is quite different for nuclei, in that their spallation cross section increases with the mass number (approximately as $A^{0.7}$). We calculated the spectrum of $^4$He nuclei accelerated at the shock and its evolution while these nuclei propagate downstream. The effect of spallation on the downstream spectra is that of depleting the low energy part of the spectrum, where the grammage is the largest. These spallation reactions result in the production of secondary $^3$He nuclei. Part of these secondary nuclei (those produced upstream of the shock) are further accelerated at the termination shock. Downstream, $^3$He nuclei get enhanced by additional fragmentation of $^4$He, while being subject themselves to spallation reactions. As a result of this chain of phenomena, we show that the total spectrum of He nuclei escaping the star cluster may be substantially hardened compared with protons, to an extent that compares well with what is required by observations, at least for a mean density $\sim 10~\rm cm^{-3}$. Larger densities result in a too large hardening, unlike what is measured. It should however be pointed out that larger mean densities become hard to justify on physical grounds, given the range of mean densities of the medium in which these star clusters are born. 

On the other hand, the effect of severe spallation losses, even for mean density of order $\sim 10~\rm cm^{-3}$, cannot be neglected for heavier nuclei. We showed that the spectra of O and Fe nuclei escaping the cluster are severely suppressed, which makes us conclude that if spallation in the bubble is the source of the discrepant hardening, then it is unlikely that star clusters can inject substantial fluxes of nuclei heavier than He in the ISM. 

We stress once more that these conclusions are sensitive to some extent to the level of clumping of the cold gas that acts as target for spallation reactions: for the typical values of parameters adopted above, all particles, at all energies, suffer the effects of spallation. On the other hand, if the density in the clumps happened to be much larger, $\gtrsim 10^4~\rm cm^{-3}$, then the number of clumps would be considerably smaller than inferred above, and the probability of encountering a clump would be, at energies $\lesssim 50$ GeV, less than unity. This would imply that some low energy nuclei might encounter a clump and suffer spallation before escaping the bubble, while others might escape unaffected by spallation.

Observationally, the best clue we may get to the clump properties may come from a combined study of HI and CO lines together with high resolution observations of the gamma ray emission. The presence of clumps should reflect in a clumpy morphology of the gamma ray emission, though averaged over the lines of sight. This may be a goal to be achieved with the upcoming generation of Cherenkov telescopes, such as CTA and ASTRI-Mini Array \citep{ASTRI:2022}.

Finally, we want to point out that the effects discussed above would equally apply to nuclei that get accelerated at the forward shock of a SNR that happens to be located inside the star cluster, when the cluster becomes old enough to experience SN explosions. The role of these SNRs for the origin of CR protons was recently investigated by \cite{Vieu-Reville:2023}, where the authors infer that the maximum energy of protons in one such sources could reach PeV. While in an isolated SNR the typical grammage traversed in the downstream region is too small to have relevant effects in terms of spallation of nuclei, this is not the case if the SNR occurs in a star cluster, where the escape of nuclei would require crossing the bubble previously excavated by the star cluster wind. Such  considerations suggest that the role of both young clusters and SNRs inside clusters should be evaluated carefully and requires a dedicated effort. 

We emphasize that a proper quantitative comparison of the predictions of these models with data requires at least two steps that at present it is premature to make: one is the inclusion of SCs of different luminosities and properties, and the other consists in accounting for SN explosions in star clusters of different ages, to the effect that these situations may lead to different spectra of escaping nuclei. We plan to develop these parts of the model in the forthcoming future.

\section*{Acknowledgements}
The authors are grateful to an anonymous referee for several useful comments and to S. Menchiari, E. Amato, N. Bucciantini, C. Evoli and O. Fornieri for fruitful discussion. GM is partially supported by the INAF Theory Grant 2022 {\it ‘‘Star Clusters As Cosmic Ray Factories''}. This work was partially funded also by the European Union – NextGenerationEU RRF M4C2 1.1 under grant PRIN-MUR 2022TJW4EJ.

\section*{Data availability}
No specific set of data was generated or analysed in support of this research. The different spectra of H and He as measured by AMS-02, were published in the relevant articles as properly cited. 

\bibliographystyle{mnras}
\bibliography{biblio} % if your bibtex file is called example.bib

% Don't change these lines
\bsp	% typesetting comment
\label{lastpage}
\end{document}